# On the timing between terrestrial gamma ray flashes, radio atmospherics, and optical lightning emission

Thomas Gjesteland[1], Nikolai Østgaard[2], Phillip Bitzer[3], and Hugh J. Christian[3]

[1]Department of Engineering Science University of Agder, Grimstad, Norway, [2]Birkeland Centre for Space Science, Department of Physics and Technology, University of Bergen, Bergen, Norway, [3]Department of Atmospheric Science, University of Alabama in Huntsville, Huntsville, Alabama, USA

**Abstract** On 25 October 2012 the Reuven Ramaty High Energy Solar Spectroscope Imager (RHESSI) and the Tropical Rainfall Measuring Mission (TRMM) satellites passed over a thunderstorm on the coast of Sri Lanka. RHESSI observed a terrestrial gamma ray flash (TGF) originating from this thunderstorm. Optical measurements of the causative lightning stroke were made by the lightning imaging sensor (LIS) on board TRMM. The World Wide Lightning Location Network (WWLLN) detected the very low frequency (VLF) radio emissions from the lightning stroke. The geolocation from WWLLN, which we also assume is the TGF source location, was in the convective core of the cloud. By using new information about both RHESSI and LIS timing accuracy, we find that the peak in the TGF light curve occurs 230 μs before the WWLLN time. Analysis of the optical signal from LIS shows that within the uncertainties, we cannot conclude which comes first: the gamma emission or the optical emission. We have also applied the new information about the LIS timing on a previously published event by Østgaard et al. (2012). Also for this event we are not able to conclude which signal comes first. More accurate instruments are needed in order to get the exact timing between the TGF and the optical signal.

**Plain Language Summary** In this paper we present two terrestrial gamma ray flash (TGF) events with observations from two different spacecraft. The Reuven Ramaty High Energy Solar Spectroscope Imager (RHESSI) and the Tropical Rainfall Measuring Mission (TRMM) satellites passed over the same thunderstorm when TGF and lightning were produced. RHESSI measured gamma ray and the lightning imaging sensor on board TRMM satellite measured optical emission from the lightning stroke. We found that the TGF (gamma rays) and the optical part of the lightning stroke were produced simultaneous to within 1.6 ms. This indicates the TGF occurs very close to the lightning stroke.

## 1. Introduction

Terrestrial gamma ray flashes (TGFs) are brief bursts of gamma radiation produced by bremsstrahlung in the Earth's atmosphere [*Fishman et al.*, 1994]. The TGFs have a typical duration of ≤1 ms [*Briggs et al.*, 2013; *Gjesteland et al.*, 2012], and they contain photons with energies up to several tens of MeV [*Smith et al.*, 2005; *Marisaldi et al.*, 2010]. Since their discovery, TGFs have been associated with thunderstorms [*Fishman et al.*, 1994] and detailed analyses of the TGF energy spectrum have shown that TGFs are produced at thundercloud altitudes. [*Dwyer and Smith*, 2005; *Carlson et al.*, 2007; *Østgaard et al.*, 2008; *Gjesteland et al.*, 2010]. Even if most studies show that TGFs are produced inside thunderstorms, it is still unknown when in the lightning process TGFs are emitted. The exact sequence of gamma rays, radio emission, and optical emission is not established. This is important for understanding the microphysics of lightning and TGFs. How TGFs are produced is one of the most important unresolved questions in the lightning process [*Dwyer and Uman*, 2014].

*Dwyer* [2012] has suggested that TGFs are produced in the ambient field of a thundercloud. In this scenario electrons are accelerated causing a cascade of electrons, which again produce photons by the bremsstrahlung process. If the photon energy is high enough (>1022 keV) photons may produce positrons by the pair production process. Photons and positrons may move backward in the electric field and create new avalanches. This process is called relativistic feedback. In this scenario TGFs can be emitted before the lightning stroke.

Another scenario for TGF production is that TGFs are produced in the strong electrical field a head of the leader tip [*Celestin and Pasko*, 2011, 2012]. Here thermal electrons are accelerated to relativistic energies and then create photons by bremsstrahlung. Also in this scenario TGFs can be produced before lightning stroke.

The electrons that produce TGFs by bremsstrahlung will produce strong radio atmospheric (sferics), which can be detected by very low frequency (VLF) receivers [*Dwyer and Cummer*, 2013]. *Cummer et al.* [2011] presented two cases where pulses in the broadband magnetic data have a strong temporal connection to the gamma ray count rates from Fermi gamma ray burst monitor (GBM). Both the magnetic field data and the Fermi data have very precise timing showing that the TGFs occur simultaneous with the sferics. Several studies have used radio data to conclude that the TGFs occur during the initial phase of intracloud lightning [*Stanley et al.*, 2006; *Shao et al.*, 2010; *Lu et al.*, 2010]. All these studies used radio emission, and not optical emission, to determine the TGF timing relative to lightning.

*Dwyer and Cummer* [2013] used a model to calculate the radio emissions from TGFs. They found that the relativistic electrons that produce TGFs can also produce powerful VLF radiation, which may be mistaken for lightning by VLF lightning location networks. *Dwyer and Cummer* [2013] also suggested, based on their model, that there may be a form of "dark lightning" which produce TGFs and VLF, but without optical light.

The first study to compare optical emission from lightning and TGF was performed by *Østgaard et al.* [2013]. *Østgaard et al.* [2013] presented simultaneous observations of TGF, VLF radio from Duke, and optical lightning measured by LIS. They concluded that the TGF occurred before the optical emission. In this paper we will show that with new information about the LIS timing we cannot conclude which one comes first: TGF or lightning.

*Connaughton et al.* [2013] compared gamma ray data from Fermi GBM and WWLLN sources. They found that TGFs are simultaneous with WWLLN to within 40 μs. They also found that short duration TGFs are more often associated with WWLLN than longer duration TGFs. *Marisaldi et al.* [2015] used the new configuration of AGILE to detect shorter TGFs. They also found that short TGFs have more WWLLN matches. *Marisaldi et al.* [2015] found the TGFs and WWLLN are simultaneous to within ±200 μs. *Mezentsev et al.* [2016] presented a detailed analysis of RHESSI and WWLLN. With the assumption that TGF is simultaneous to the WWLLN, *Mezentsev et al.* [2016] concluded that the RHESSI clock has a varying systematic delay but the uncertainty is less than 100 μs.

In this paper we will present two TGFs that were observed by RHESSI (gamma ray), LIS (optical), and WWLLN (radio). In section 2 we present the data and the observations. In section 3 we present the timing uncertainties of the TGF measured by RHESSI, sferics measured by WWLLN, and optical measured by LIS. One of the TGF is a new event not published before. We have also reanalyzed the timing of a previously published TGF by *Østgaard et al.* [2013] . We have performed a statistical study of the time difference between WWLLN and LIS which is presented in section 3.1. The discussion and conclusion are presented in sections 4 and 5.

## 2. Data and Observations

RHESSI is a small NASA explorer that was designed to study solar flares. It consists of nine germanium detectors that measure photon energies up to 17 MeV [*Grefenstette et al.*, 2009]. RHESSI time tags each photon with $2^{-20} = 0.9537$ μs resolution and telemeters all data to ground. The RHESSI clock offset with respect to UTC is estimated to be $-1.808 \pm 0.050$ ms for the period 5 August 2005 to 21 October 2013 [*Mezentsev et al.*, 2016].

The Tropical Rainfall Measuring Mission (TRMM) is an observatory containing instruments to study weather and lightning. It contains the precipitation radar (PR) , the TRMM microwave imager, and the visible and infrared scanner (VIRS). Lighting imaging sensor (LIS) is a lightning detector that measures optical emissions from the atomic oxygen line at 777.4 nm. The detector is a 128 by 128 pixelated charge-coupled device that records 559 frames per second. This corresponds to either 1.495 or 2.014 ms integration time per frame. The data provided are events, groups, and flashes. An event is one illuminated pixel. Adjacent events in the same frame are a group and groups occurring within 330 ms and within 5 km are a flash. Based on ground-based lightning detection network over North America, LIS is estimated to detect 80% of all lightning [*Bitzer et al.*, 2016]. A full description of the timing of LIS can be found at *Bitzer and Christian* [2015].

Figure 1 shows a map of southern India and Sri Lanka where the TGF was detected. The RHESSI subsatellite point is shown as a red triangle and the black circle is the distance of 1000 km from the subsatellite point. The grey dots show the LIS instruments field of view (FOV) for the orbit. The light blue color is the LIS FOV at the time of the TGF. The black dots are the locations of lightning measured by LIS. The orange asterisk shows the location of the LIS lightning, which is simultaneous with the TGF. The TGF occurred at the west coast of Sri Lanka. The distance between the lightning and the RHESSI subsatellite point is 297 km. The precipitation radar (PR) and the visible infrared radiometer (VIRS) show that the TGF comes from the convective core of this thundercloud.

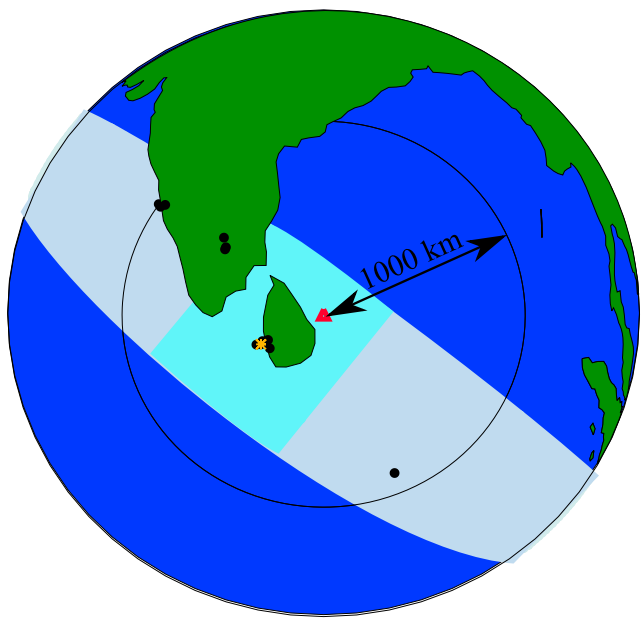

**Figure 1.** A map of the location of the TGF on 25 October 2012. The LIS orbits field of view in grey and the field of view at the time of the TGF in blue. The red triangle is the location of RHESSI. Black dots are WWLLN locations, and the asterisk is the location of the TGF.

Figure 2 shows a 20 ms window with reference time 25 October 2012 13:27:13.221 UT. The RHESSI counts are shown as blue dots with a logarithmic energy scale on the y axis. The systematic 1.808 ms delay has been taken into account as well as the light travel time from the lightning to RHESSI. The TGF is assumed to occur at the same location as the lightning measured by WWLLN. Indicated with a red line is the time of a WWLLN lightning from the same location as the LIS lightning measurement. A lightning was detected by WWLLN at occurred at 13.221327 s. The figure shows that the WWLLN measurement is simultaneous to the RHESSI TGF.

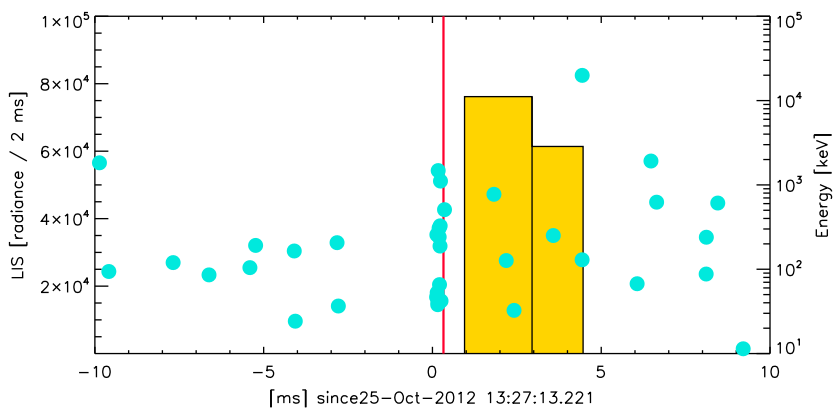

**Figure 2.** Blue dots are the times of RHESSI photons plotted against their energy. An offset of 1.808 ms is added to the RHESSI data, and the times are shifted back to source location via light travel time correction. The RHESSI clock has 0.050 ms uncertainty. WWLLN detects a lightning stroke indicated by the red line at time 25 October 2012 13:27:13.221327 UT (+0.327 ms with respect to reference time). The LIS measurements are shown in yellow. The first time tag (1 ms prior to the end of the frame) is at +1.98 ms with respect to reference time.

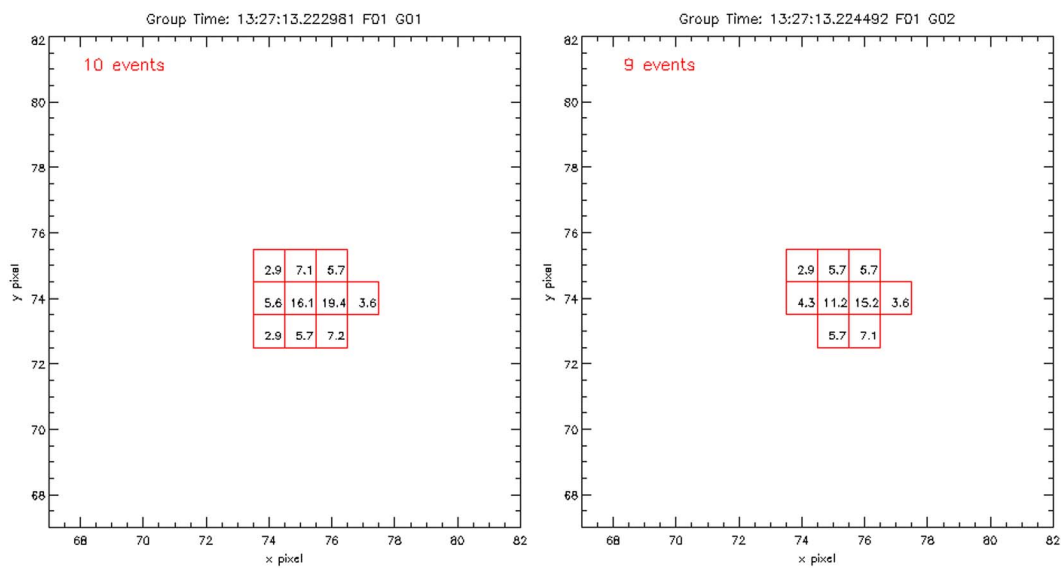

**Figure 3.** The axis is pixel addresses for the events in the two frames in Figure 2. The number inside each pixel is the measured radiance in J/sr/m$^2$/μs over 2.014 and 1.495 ms, respectively. (left) The first frame and (right) the second frame. In both frames the brightest events are located close to the center (pixel [74, 75] and pixel [74, 76])

However, the peak of a Gauss fit to the TGF light curve is 230 μs before the WWLLN measurement. The duration of the TGF is 0.332 ms. The yellow histogram shows the LIS measurements in radiance. LIS measured two groups. The first group had time tag of 13.222981 s which corresponds to 1.961 ms after the reference time in Figure 2. The next group had time tag of 13.224492 s, which corresponds to 3.492 ms after the reference time. The LIS time tag is 1 ms prior to the end of the frame and the integration time for each frame is either 1.495 ms or 2.014 ms [*Bitzer and Christian*, 2015].

Figure 3 shows the pixel distribution for each of the LIS bins in Figure 2. On each pixel the radiance value is given. The first group contains 10 events (events = recording in one pixel) where the two brightest events, which are located in center (pixel [74, 75] and pixel [74, 76]) have radiance of 16.1 J/sr/m$^2$/μs and 19.4 J/sr/m$^2$/μs, respectively. In the next group the same two pixels are still the brightest with radiance of 11.2 J/sr/m$^2$/μs and 15.2 J/sr/m$^2$/μs. These observations indicate that LIS measures a lightning in the center and that the surrounding events are illuminated due to scattered light from the cloud. We also see that 59% of the energy arrives in the first group. Assuming an optical pulse width of 400 μs and an optical rise time of 100 μs, the optical pulse started at 13.2228 s which correspond to 1.8 ms after the reference time in Figure 2. With this assumption the LIS lightning time is 1.7 ms after the TGF (TGF-time at source: 13.221097 s).

Figure 4 shows a 20 ms window of the second TGF event with reference time 27 October 2006 04:56:03:000 UT. The RHESSI counts are shown as blue dots with a logarithmic energy scale on the *y* axis. The systematic 1.808 ms delay has been taken into account as well as the light travel time from the lightning to RHESSI. The black curve is the Duke VLF measurements. This event was also presented in *Østgaard et al.* [2013]. The black histogram is the LIS measurements as presented in *Østgaard et al.* [2013]. The yellow histogram is the LIS measurements when the time corrections from *Bitzer and Christian* [2015] is accounted for.

### 3. Timing Uncertainties

The timing of the RHESSI clock has been suggested to have a constant offset. In a comparison between RHESSI TGFs and WWLLN *Mezentsev et al.* [2016] found that the RHESSI clock has a systematic 1.808 ms offset and a random uncertainty of 0.050 ms. This offset is valid for the period 5 August 2005 to 21 October 2013.

The timing of LIS has been addressed by *Bitzer and Christian* [2015]. The timing of LIS is 1 ms prior to the end of the frame. The LIS time tag was set before the TRMM was boosted from 350 km altitude to 402 km altitude, and this distance must be accounted for. *Bitzer and Christian* [2015] also show that the time of LIS assumes that the source is at nadir. All off nadir events must be corrected for. *Bitzer and Christian* [2015] presents this equation to correct the LIS time:

$$t_s = \frac{r_{pre}}{c} - \frac{r_{post}}{c} - \frac{r_{off}}{c} + t_{LIS}, \qquad (1)$$

where $r_{pre}/c$ = 1.27 ms and $r_{post}/c$ = 1.30 ms. $r_{off}/c$ varies with off nadir pixels, and $t_{LIS}$ is the time provided by the LIS data.

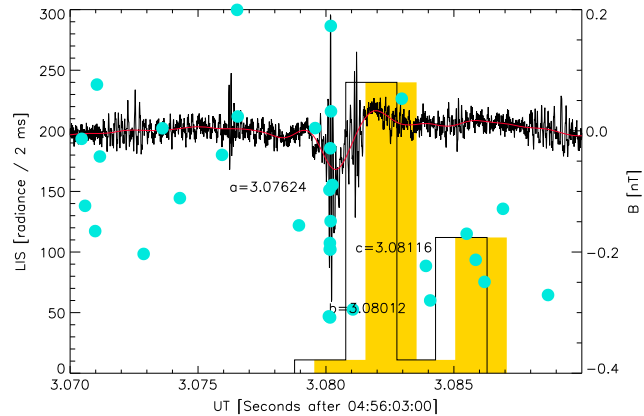

**Figure 4.** The timing of the TGF event on 27 October 2006. The yellow histogram is the LIS measurements when the time corrections from *Bitzer and Christian* [2015] is accounted for. The black histogram is the LIS times presented in *Østgaard et al.* [2013]. Duke VLF signals are shown in black overlaid with a smoothed red line. The TGF is simultaneous with the first Duke radio pulse. The second Duke radio pulse is not simultaneous with the optical LIS measurement.

For the events presented in this paper the brightest LIS event has pixel address [74, 76], which is very close to the nadir point. Adjusting for the altitude boost corresponds to a 30 µs correction.

In *Østgaard et al.* [2013] we used the end of the frame as the LIS time. According to *Bitzer and Christian* [2015], this is wrong. To get the correct timing for this event, we must first correct the times with 1 ms prior to the end of the frame. This LIS measurement is on *x* pixel 77 and *y* pixel 29. This is ∼30 pixel off nadir, and it corresponds to a time delay of 100–200 µs. When adding the delay due to postboost, 30 µs, and the 1 ms time change due to center of the bin (+1 ms) the total time shift in LIS is 770–870 µs. In *Østgaard et al.* [2013] 1.9 ms was used as the constant RHESSI offset. This was calculated using events before and after 5 August 2005. New corrections to the RHESSI clock is 1.808 ± 0.050 ms [*Mezentsev et al.*, 2016]. The time of RHESSI is therefore 0.092 ms earlier than reported in *Østgaard et al.* [2013]. Including all these corrections, the time between RHESSI and LIS is not 0.4 ms as reported by *Østgaard et al.* [2013], but 1.26–1.36 ms (see Figure 4).

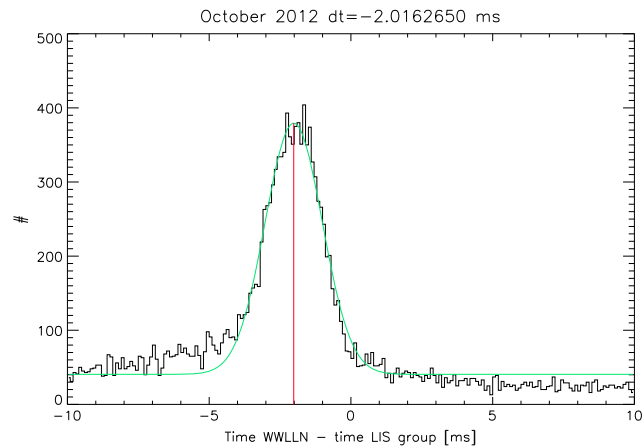

**Figure 5.** The time between WWLLN lightning and the closest group time measured by LIS. The green curve is a Gaussian fit. The centre of the Gaussian fit is marked by the vertical red line at −2.01 ms. The standard deviation is 1.3 ms. In the figure 16,946 measured by both WWLLN and LIS from October 2012 are included. In this data are only lightning which the time between LIS group and WWLLN was less than 10 ms and the distance between the two lightning locations was less than 45 km.

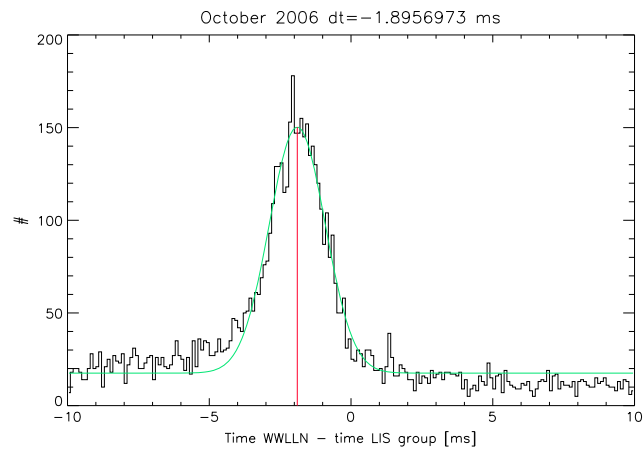

**Figure 6.** The time between WWLLN lightning and the closest group time measured by LIS. The green curve is a Gaussian fit. The centre of the Gaussian fit is marked by the vertical red line at −1.90 ms. The standard deviation is 1.4 ms. In the figure 6912 lightning measured by both WWLLN and LIS from October 2006 are shown. In this data are only lightning where the time between LIS group and WWLLN is less than 10 ms and the distance between the two lightning locations is less than 45 km.

We have here presented two cases where the TGF peak time was before the LIS times. The peak time of one TGF was 1.26–1.36 ms before the LIS measurement and another event had the peak time of the TGF 1.7 ms before the LIS measurement. In the next section we present an analysis of timing between LIS and WWLLN. This analysis shows that the LIS clock has a systematic delay of 1.5 ms with an uncertainty of 1.4 ms. Adding this delay, we find that the TGFs occurred 0.24 ± 1.4 ms before and 0.2 ± 1.4 ms after the optical for our two events. Due to the uncertainty in the LIS instruments clock, we cannot conclude that TGF comes before the optical emission of the lightning.

### 3.1. Statistical Timing Between WWLLN and LIS

Figures 5 and 6 show histograms of the time difference between WWLLN and LIS groups for October 2012 and October 2006. The method we have used to find lightning that are measured by both LIS and WWLLN is the following: For each LIS group we calculate $\Delta t$ for the WWLLN lightning that occurs closest in time and we include only events where $|\Delta t| < 10$ ms. We only include lightning where the difference in position between the LIS group and WWLLN is <45 km, which is the WWLLN uncertainty. With these constraints we limit the data set to only contain lightning that are observed by both WWLLN and LIS.

Both Figures 5 and 6 show $t_w − t_S$, where $t_w$ is the time reported by WWLLN and $t_S$ is the time of the LIS group at source calculated by equation (1).

Figure 5 contains 1 month of data from October 2012. The vertical red lines in Figure 5 is the center of the Gaussian fit to the distribution, and this is our estimation of $\Delta t$. Figure 6 is a similar plot with data from October 2006. The estimated $\Delta t$ is −2.01 ± 1.3 ms in October 2012 and −1.90 ± 1.4 ms in October 2006. We will use 1.4 ms as the uncertainty for the time between WWLLN and LIS.

A similar study has been preformed by *Bitzer et al.* [2016]. They compared LIS group and lightning data from Earth Networks Total Lightning Network (ENTLN). They found that the ENTLN time tag is on average 1.7 ms before the LIS group time [*Bitzer et al.*, 2016, Figure 8].

## 4. Discussion

The timing between VLF radio from WWLLN and optical signal has been studied by *Lay et al.* [2007]. They used optical data form the photo diode detector (PDD) on the Fast On-Orbit Recording of Transient Events (FORTE) satellite and compared the time measurements with WWLLN measurements. PDD has a circular field of view with radius 1200 km and is a nonimaging instrument. It triggers on rising optical signal intensity. Its sample step is 15 μs. *Lay et al.* [2007] used the WWLLN location as the true location of the lightning and corrected the PDD timing measurements with the time of flight from ground to satellite. Figure 1 in *Lay et al.* [2007] shows a histogram of WWLLN time-FORTE PDD time. The peak is at −0.25 ms, which means that the optical signal measured by FORTE PDD is later than the WWLLN time. *Lay et al.* [2007] also presented a superposed epoch accumulation of the PDD waveform and the WWLLN time [*Lay et al.*, 2007, Figure 3]. They found that there is a 0.3–0.4 ms delay in the optical peak with respect to the WWLLN time.

According to *Suszcynsky et al.* [2000], this delay is caused by two reasons. First, there might be a delay between the radio emission and the optical emission in lightning. Second, the photons are scattered as it propagates through the cloud. This causes a time delay for the optical signal. *Suszcynsky et al.* [2000] used VHF data measured by FORTE and the FORTE PDD to estimate that the mean delay between VHF and PDD was 243 μs. The transient delay had mean of 105 μs, and the mean delay due to scattering was 138 μs, which corresponds to 41 km additional path length as the photons propagate through the atmosphere.

The analysis of the gamma ray data and WWLLN data shows that TGF and WWLLN occur close in time. The analysis of optical measurements from LIS shows that TGF occur prior to the optical emission. From the literature we found that 0.3–0.4 ms could be explained by a physical delay between radio emission, and a part caused by scattering of light as it propagates through the atmosphere.

Our statistically analysis of WWLLN and LIS shows that the LIS time are 1.9–2.0 ± 1.4 ms later than WWLLN. Since 0.3–0.4 ms can be explained as physical delay and scattering of light as it propagates through the atmosphere, we suggest that the LIS clock has a systematic delay of 1.5 ± 1.4 ms.

The delay between LIS and radio data is also found by *Bitzer et al.* [2016] who compared LIS groups and lightning data from ENTLN. They found a delay of 1.7 ms.

## 5. Conclusion

We have analyzed two events where we have both radio (WWLLN), optical (LIS), and gamma ray (RHESSI) observations. Based on radio and optical measurements, we can determine the location of the lightning to within 45 km. The timing of radio and gamma have uncertainties of 15 μs and 50 μs, respectively. A statistical comparison between WWLLN and LIS indicate a 1.5 ± 1.4 ms systematic delay of the LIS clock. With these uncertainties we cannot determine the sequence of events, only that the TGF and the optical signal are simultaneous to within ±1.6 ms.

**Acknowledgments**
This study was supported by the Research Council of Norway under contracts 184790/V30 and 197638/V30. We thank the RHESSI team for the use of RHESSI raw data and software. The LIS/OTD High Resolution Annual Climatology (HRAC) data were produced by the LIS/OTD Science Team (principal investigator, Hugh J. Christian). We thank the institutions contributing to WWLLN (http://wwlln.net/). We thank Steven A. Cummer for the Duke VLF data shown in Figure 4. The RHESSI data are availible at https://hesperia.gsfc.nasa.gov/rhessi3 The LIS data are availible at https://lightning.nsstc.nasa.gov/data/data_lis.html.